\documentclass[reprint,superscriptaddress,amsmath,amssymb,aps,twocolumn,pra]{revtex4-1}

\usepackage{color,graphicx,amsfonts}
\usepackage{multirow,amsbsy}
\usepackage{verbatim}
\usepackage{bm}
\usepackage{bbold}
\usepackage{soul}
\usepackage{hyperref}

\newcommand{\bra}[1]{\ensuremath{\left\langle #1\right|}}
\newcommand{\ket}[1]{\ensuremath{\left|#1\right\rangle}}

\newcommand{\sminus}{\rule[0.6ex]{0.4em}{0.08ex}}

\begin{document}

\title{Nonunitary Preparation of Nontrivial States from Trivial Regimes in Two-Dimensional Topological Insulators}

\author{Qin-Qin Wang}
\affiliation{School of Physics, Hefei University of Technology, Hefei 230026, China}
\affiliation{Laboratory of Quantum Information, University of Science and Technology of China, Hefei 230026, China}
\affiliation{Anhui Province Key Laboratory of Quantum Network, University of Science and Technology of China, Hefei 230026, China}
\affiliation{CAS Center for Excellence in Quantum Information and Quantum Physics, University of Science and Technology of China, Hefei 230026, China}

\author{Xiao-Ye Xu}
\email{xuxiaoye@ustc.edu.cn}
\author{Chuan-Feng Li}
\email{cfli@ustc.edu.cn}
\author{Guang-Can Guo}
\affiliation{Laboratory of Quantum Information, University of Science and Technology of China, Hefei 230026, China}
\affiliation{Anhui Province Key Laboratory of Quantum Network, University of Science and Technology of China, Hefei 230026, China}
\affiliation{CAS Center for Excellence in Quantum Information and Quantum Physics, University of Science and Technology of China, Hefei 230026, China}
\affiliation{Hefei National Laboratory, University of Science and Technology of China, Hefei 230088, China}

\begin{abstract}

While remarkable progress has been achieved in engineering nontrivial Hamiltonians across a wide range of physical platforms, preparing their corresponding nontrivial ground states remains a major experimental challenge.
The commonly used strategy for state preparation relies on adiabatic protocols.
However, when a trivial initial state is unitarily driven toward nontrivial regimes, the dynamics must cross gap-closing critical points, rendering the process intrinsically nonadiabatic, and the state remains topologically trivial.
Here, we present a nonunitary method for dynamically preparing nontrivial states in two-dimensional topological insulators.
By introducing dephasing noise into a slowly driven unitary evolution, we demonstrate that the topological number of the resulting dephased states can coincide with that of the target nontrivial Hamiltonian.
This nearly adiabatic nonunitary state-preparation protocol provides a powerful alternative to conventional adiabatic approaches for accessing topological states.

\end{abstract}

\maketitle

\section{Introduction}
Engineering quantum systems into low-temperature states and exploring the underlying remarkable properties of topologically nontrivial phases are subjects of intense current interest\,\cite{Hasan2010,Chiu2016,Ren2016}.
A widely pursued strategy for such state preparation is the adiabatic process\,\cite{Albash2018}, in which system parameters are varied slowly to suppress nonadiabatic excitations so that the state follows the instantaneous ground state at all times.
When the adiabatic process approaches gapless topological phase boundaries, maintaining adiabaticity requires increasingly long evolution times\,\cite{Odelin2019}, which typically exceed the coherence times of realistic experimental platforms.
To accommodate finite coherence times, significant efforts have been devoted to accelerating state-preparation protocols while preserving adiabatic performance\,\cite{Campo2013,Zhou2020,Takahashi2024,Jernej2025,Morawetz2025}.

However, when unitarily evolving a topologically trivial ground state into a nontrivial regime, the time-dependent driven Hamiltonian necessarily crosses a topological phase boundary.
As a result, the evolution becomes nonadiabatic due to unavoidable excitations generated near the gap-closing points\,\cite{Zener1932}.
Moreover, coherent dynamics--being local unitary transformations--cannot alter the trivial topological character of the state\,\cite{Chen2010,McGinley2018,Reid2022}; in particular for two-dimensional (2D) topological insulators, its Chern number has been shown to remain zero throughout the evolution\,\cite{Caio2015,McGinley2019}.
To circumvent the limitations imposed by gap closing and enable the preparation of topologically nontrivial states, alternative strategies have been proposed, including periodically driven protocols that introduce nontrivial Floquet Hamiltonians with adiabatic driving and enable topological state preparation\,\cite{Luca2015,Bandyopadhyay2019}, as well as dissipative schemes\,\cite{Budich2015a,Bandyopadhyay2020a,Liu2021,Alexander2024,Ding2024,Pocklington2025,Li2025,Zhan2026} in which the desired topological states are encoded as the unique steady state of a quantum master equation or obtained by coupling the system to a conjugate copy and projecting onto the target state\,\cite{Barbarino2020}.

In this work, we propose a nonunitary protocol for preparing topologically nontrivial states, illustrated using the 2D Qi-Wu-Zhang model\,\cite{Qi2006}.
The system is initialized in a topologically trivial ground state, and the Hamiltonian parameters are slowly ramped into the nontrivial regime.
By introducing dephasing noise into the unitary dynamics, the coherent superposition between the ground and excited states is fully suppressed.
We show that, in the weak-excitation limit, the resulting dephased state becomes topologically nontrivial.
And its topological number is always consistent with that of the driven Hamiltonian under this nearly adiabatic nonunitary evolution.
Here, the topological properties of the dephased states are characterized by the Uhlmann number\,\cite{Viyuela2014,Viyuela2015}, a mixed-state generalization of the Chern number, together with an equivalent topological index defined from the spectrum of the holonomy matrix\,\cite{Huang2014}.

This paper is organized as follows: In Sec.\,\ref{sec2}, we will introduce two equivalent mixed-state topological invariants, and detail the nonunitary state-preparation protocol in an exemplary 2D topological insulator model.
In Sec.\,\ref{sec3}, we will present numerical analyses of our nontrivial state preparation protocol and examine the effects of the dephasing noise strength and the parameter ramp velocity.
Finally, Sec.\,\ref{sec4} summarizes our results and provides a discussion.

\newpage

\section{\label{sec2}THEORETICAL IDEA}
In this section, we outline the theoretical framework underlying the nonunitary protocol for preparing topologically nontrivial states in 2D topological insulators.

\subsection{Mixed-state topological invariants}

For pure ground states in 2D topological insulators, the topological properties are characterized by means of the Chern number, which can be written as\,\cite{Abanin2013,Wang_2019,Chen2025}:
\begin{equation}\label{eq.ch}
    n_{\mathrm{C}} = \frac{1}{2\pi} \oint dk_x\frac{d\Phi_{\mathrm{B}}(k_x)}{dk_x},
\end{equation}
where $\Phi_{\mathrm{B}}(k_x)=i\oint dk_y \langle \mathbf{n}_\mathbf{k}|\partial_{k_y}|\mathbf{n}_\mathbf{k}\rangle$ is the Berry geometric phase accumulated along a closed loop in the $k_y$ direction at fixed $k_x$, and $\mathbf{n}_\mathbf{k}$ is the eigenvector.
The Chern number counts the number of $2\pi$-discontinuous jumps of $\Phi_{\mathrm{B}}(k_x)$ as $k_x$ traverses the Brillouin zone and is therefore quantized to integer values.
Conversely, if $\Phi_{\mathrm{B}}(k_x)$ varies continuously and does not sweep the full interval $[-\pi,\pi]$ along the $k_x$ loop, then $n_{\mathrm{C}}=0$ and the corresponding state is topologically trivial.
If $\Phi_{\mathrm{B}}(k_x)$ exhibits $2\pi$-discontinuous jumps, the Chern number is nonzero, and the state is topologically nontrivial.

To characterize the topology of general density matrices $\rho_{\mathbf{k}}$ arising in nonunitary dynamics, we introduce the topological Uhlmann number\,\cite{Viyuela2014,Viyuela2015}, obtained by replacing the Berry geometric phase in Eq.\,\eqref{eq.ch} with the Uhlmann geometric phase\,\cite{Viyuela2014a}.
The Uhlmann geometric phase is formulated in terms of an amplitude $\omega_{\mathbf{k}}$ satisfying
$\rho_{\mathbf{k}}=\omega_{\mathbf{k}}\omega_{\mathbf{k}}^{\dagger}$\,\cite{Viyuela2014a,Budich2015,Morachis2021,Zhang2021,Xin2025}.
The amplitude $\omega_{\mathbf{k}}=\sqrt{\rho_{\mathbf{k}}}\,U$ with $U$ is unitary matrix, which can be viewed as a purification of the mixed state.
Under Uhlmann parallel transport\,\cite{Uhlmann1986} along a closed trajectory parametrized by $\mathbf{k}$, the amplitude acquires a unitary matrix,
$\omega_{\mathbf{k}(1)}=\omega_{\mathbf{k}(0)}V$,
where $\omega_{\mathbf{k}(0)}$ and $\omega_{\mathbf{k}(1)}$ denote the initial and final points of the transport.
Here, the Uhlmann holonomy $V=\mathcal{P}e^{\oint A_{\mathrm{U}}}U_0$ is determined by the Uhlmann connection $A_{\mathrm{U}}$, with $\mathcal{P}$ denoting path ordering and $U_0$ is the choice of gauge taken for the initial amplitude $\omega_{\mathbf{k}(0)}$.
The Uhlmann geometric phase is defined as the relative phase between the initial and final amplitudes,
$\Phi_{\mathrm{U}}=\arg(\omega_{\mathbf{k}(0)},\omega_{\mathbf{k}(1)})=\arg \text{Tr}[\omega^{\dagger}_{\mathbf{k}(0)}\omega_{\mathbf{k}(1)}]$.
This gauge-independent geometric phase can be expressed as:
\begin{equation}
    \Phi_\mathrm{U} 
    =  \text{arg} (\text{Tr}[\rho_{\textbf{k}(0)} \mathcal{P}\text{e}^{\oint A_\mathrm{U}}]).
\end{equation}
The Uhlmann connection is given by:
\begin{align}
    A_\mathrm{U} = \sum_{u,i,j}\frac{|\phi^{i}_{\mathbf{k}}\rangle\langle\phi^{i}_{\mathbf{k}}|[\partial_{u}\sqrt{\rho_{\mathbf{k}}},\sqrt{\rho_{\mathbf{k}}}]|\phi^{j}_{\mathbf{k}}\rangle\langle\phi^{j}_{\mathbf{k}}|}{p^{i}_{\mathbf{k}}+p^{j}_{\mathbf{k}}} dk_u,
\end{align}
where the mixed state is expressed in its spectral decomposition $\rho_\mathbf{k}=\sum_{i=\pm} p_\mathbf{k}^{i}|\phi_\mathbf{k}^{i}\rangle\langle \phi_\mathbf{k}^{i}|$.

After replacing the Berry geometric phase in the definition of Chern number in Eq.\,\eqref{eq.ch} with the Uhlmann geometric phase, the Uhlmann number can be expressed as\,\cite{Viyuela2014}:
\begin{equation}\label{eq.uh}
    n_{\mathrm{U}} = \frac{1}{2\pi} \oint dk_x\frac{d\Phi_{\mathrm{U}}(k_x)}{dk_x}.
\end{equation}
Analogous to the Chern number, the Uhlmann number characterizes the winding number of $\Phi_{\mathrm{U}}(k_x)$ along the $k_x$ loop and takes quantized integer values.
In the zero-temperature limit, the Uhlmann geometric phase reduces to the Berry geometric phase (modulo $2\pi$), and consequently the Uhlmann number reduces to the conventional Chern number\,\cite{Viyuela2015}.
Equivalently, the topological properties of mixed states can be analyzed through the spectrum of the holonomy matrix $M=\omega^{\dagger}_{\mathbf{k}(0)}\omega_{\mathbf{k}(1)}$\,\cite{Huang2014}.
Its eigenvalues are generally complex and can be written in polar form as $z_{\pm}=\lambda_{\pm}e^{i\mu_{\pm}}$ for two-band models.
$\lambda_{\pm}$ are amplitudes with larger/smaller values, and $\mu_{\pm}$ denote the corresponding phases as the geometric phases of the cyclic path mapped to the matrix $M$. 
The two phases are opposite to each other, i.e., $\mu_{+}=-\mu_{-}$.
When the amplitude spectra $\lambda_{\pm}(k_x)$ remain gapped, a topological index of mixed states $\tilde{n}_{\mathrm{U}}$ can be defined from the winding number of the phases $\mu_{\pm}(k_x)$ as $k_x$ traverses the loop.
A change in $\tilde{n}_{\mathrm{U}}$ is necessarily accompanied by a closing of the amplitude gap, in close analogy with gap closings of energy bands at zero temperature.
In the following, we use the two equivalent Uhlmann invariants $n_{\mathrm{U}}$ and $\tilde{n}_{\mathrm{U}}$ to characterize the topological properties of the mixed states generated by the nonunitary dynamics.

\subsection{Nonunitary state-preparation protocol}

\begin{figure*}
  \centering
  \includegraphics[width=0.98\textwidth]{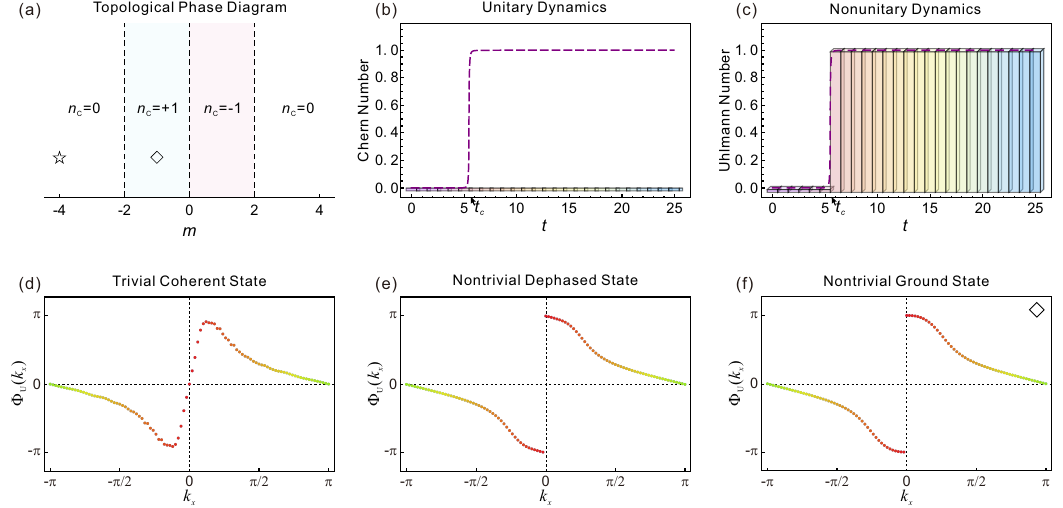}
  \caption{\textbf{Nonunitary preparation of topologically nontrivial states.}
  (a) Topological phase diagram of the Qi-Wu-Zhang model.
  White, blue, and red regions correspond to topological phases with lower-band Chern numbers $n_{\mathrm{C}}=0$, $+1$, and $-1$, respectively.
  Black dashed lines indicate the topological phase boundaries associated with energy-gap closings at $\mathbf{k}_c\in\{0,\pi\}$.
  The Hamiltonian parameter is ramped from an initially trivial regime (star symbol) to a final nontrivial regime (diamond symbol) along the trajectory $m(t)=m_i+(m_f-m_i)(1-e^{-vt})$.
  Parameters are $m_i=-4$, $m_f=-1$, and $v=0.2$.
  (b) Chern number of the coherently evolved state (bar chart) compared with that of the instantaneous Hamiltonian (purple dotted line).
  (c) Uhlmann number of the instantaneously dephased state (bar chart) compared with that of the instantaneous Hamiltonian (purple dotted line). 
  The dephasing rate is set to $\gamma_{\mathbf{k}}=2$.
  (d–f) Uhlmann geometric phase of the coherently evolved final state, the dephased final state, and the ground state of the final nontrivial Hamiltonian at $t=25$, respectively, as functions of $k_x$.
  The topological number of the evolving state in (b,\,c) is plotted at discrete times. 
  The momentum space $k_x\in[-\pi,\pi]$ in (d-f) is discretized into 101 equally spaced points. 
  }\label{fig.results1}
\end{figure*}

We study a 2D topological insulator known as the Qi-Wu-Zhang model\,\cite{Qi2006}, which has been widely employed to investigate the quantum Hall effect and can be physically realized\,\cite{Liu2008}.
The Bloch Hamiltonian of this two-band model can be written as:
\begin{equation}
   H_{\mathbf{k}}(m) =  \mathbf{d}_{\mathbf{k}}(m)\cdot\boldsymbol{\sigma} = \mathbf{E}_{\mathbf{k}}(m) \mathbf{n}_{\mathbf{k}}(m)\cdot\boldsymbol{\sigma},
\end{equation}
where $\boldsymbol{\sigma}=(\sigma_x,\,\sigma_y,\,\sigma_z)$ are the Pauli matrices and $\mathbf{d}_{\mathbf{k}}(m) = (\sin k_x,\,\sin k_y,\,m+\cos k_x +\cos k_y)$ is a real three-component vector.
The energy spectrum is given by $\pm E_{\mathbf{k}}=\pm|\mathbf{d}_{\mathbf{k}}|$, and the corresponding eigenvector $\mathbf{n}_{\mathbf{k}}=\mathbf{d}_{\mathbf{k}}/|\mathbf{d}_{\mathbf{k}}|$.
The first Brillouin zone is given by $k_x,k_y\in[-\pi,\pi]$.
For this model, the Chern number of the lower band of $H_{\mathbf{k}}(m)$ is $n_{C}=-\text{sgn}(m)$ for $0<|m|<2$ and $n_{C}=0$ for otherwise.
The resulting topological phase diagram as a function of $m$ is shown in Fig.\,\ref{fig.results1}(a).

We consider a dynamical state-preparation protocol in which the system is initially prepared in the ground state of a topologically trivial Hamiltonian $H(m_i)$, and the parameter $m(t)$ is slowly ramped to a target nontrivial Hamiltonian $H(m_f)$ along the trajectory $m(t)=m_i+(m_f-m_i)(1-e^{-vt})$.
When the ramp process crosses a gapless phase boundary, excitations are inevitably created near the gap-closing points, irrespective of how small the ramp velocity $v$ is, leading to the coherent superposition of ground and excited states\,\cite{Zener1932}.
Crucially, such coherent dynamics cannot alter the topological character of the state: the Chern number of the instantaneous state remains trivial, $n_{\mathrm{C}}(t)\equiv0$, throughout the evolution\,\cite{Caio2015}.
Consequently, the preparation of a topologically nontrivial state from a trivial initial state is prohibited under purely unitary dynamics.

\begin{figure*}
  \centering
  \includegraphics[width=0.98\textwidth]{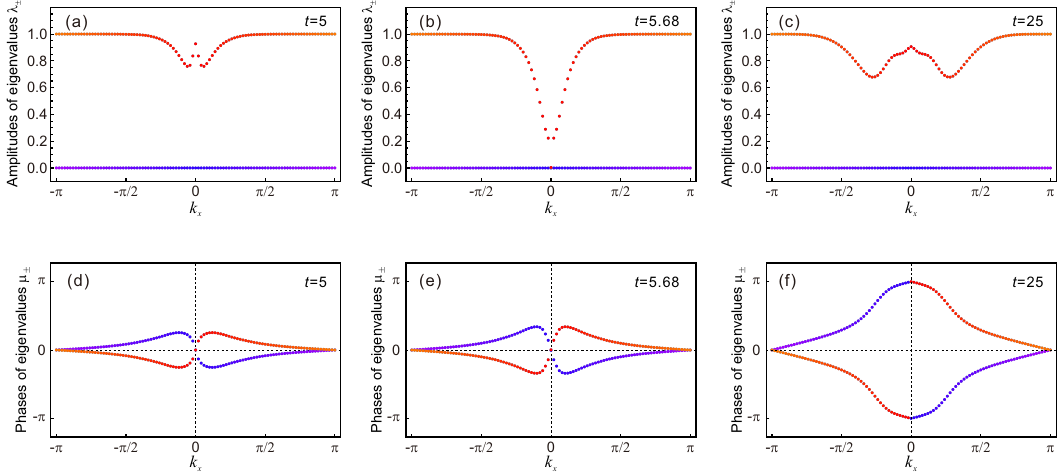}
  \caption{\textbf{Topological phase transition in parameter ramp.}
  Top panels (a-c) show the amplitude spectra of the holonomy eigenvalues $\lambda_{\pm}(k_x)$ for the evolving dephased state at $t=5$, $5.68$, and $25$, respectively.
  Bottom panels (d-f) display the corresponding phases of the holonomy eigenvalues $\mu_{\pm}(k_x)$ as a function of $k_x$.
  Red and blue dots represent the complex eigenvalues $z_{\pm}$ associated with the larger and smaller amplitudes, respectively.
  Parameters are $m_i=-4$, $m_f=-1$, $v=0.2$ and $\gamma_{\mathbf{k}}=2.0$ for all panels. }\label{fig.results2}
\end{figure*}

To overcome this limitation, we introduce a purely dephasing mechanism that completely suppresses the coherent superposition between the ground and excited states.
The resulting nonunitary dynamics is described by the master equation\,\cite{Ying2016}:
\begin{equation}\label{eq.dephasing}
    \dot{\rho}_{\mathbf{k}} = -\text{i}[H_{\mathbf{k}},\rho_\mathbf{k}] + \gamma_{\mathbf{k}}[\tilde{\sigma}^z_\mathbf{k} \rho_\mathbf{k} \tilde{\sigma}^z_\mathbf{k} - \rho_\mathbf{k}],
\end{equation}
where the overdot denotes the time derivative, $\tilde{\sigma}^z_\mathbf{k} (t) = \mathbf{n}_{\mathbf{k}}(t)\cdot\boldsymbol{\sigma}$
is the Pauli operator defined in the instantaneous eigenbasis of $H_{\mathbf{k}}(m(t))$, and $\gamma_{\mathbf{k}}$ is the dephasing rate, which depends on momentum.
Under this nonunitary evolution in the strong-dephasing regime, the dephased state becomes fully diagonal in the instantaneous eigenbasis and takes the form:
\begin{equation}
    \rho^{\text{de}}_{\mathbf{k}} = \frac{1}{2}[\mathbb{1}+r_\mathbf{k}\tilde{\sigma}^z_\mathbf{k}].
\end{equation}
Here, $r_{\mathbf k} = 2P_{\mathrm{LZ}}^{\mathbf k}-1$.
For a slow parameter ramp $v\to 0$, the Landau--Zener excitation probability $P_{\mathrm{LZ}}^{\mathbf k}$ is exponentially small away from the gap-closing
momentum $\mathbf k_c$, and hence $r_\mathbf{k}$ approaches -1.
And $r_\mathbf{k}\in [-1,1]$ in the narrow excitation region near $\mathbf{k}_c$. 
The populations of the upper and lower bands, $(1\pm r_\mathbf{k})/2$, are conserved under the pure dephasing mechanism, whereas the off-diagonal elements of $\rho_\mathbf{k}$ in the eigenstate of the instantaneous Hamiltonian are completely suppressed.
As a result, the dephased state is mixed, with purity $(1+|r_{\mathbf{k}}|^2)/2$, rendering the conventional Chern number inapplicable and motivating the adoption of mixed-state topological invariants in Eq.\,\eqref{eq.uh}.

In regimes where the amplitude spectrum is gapped ($\Delta_{\lambda}(k_x)=\lambda_{+}(k_x)-\lambda_{-}(k_x)>0$ for all $k_x\in[-\pi,\pi]$), the mixed-state topological invariants $\tilde{n}_{\mathrm{U}}$ and $n_{\mathrm{U}}$ are determined by the phase $\mu_{+}(k_{x_c})$, associated with the larger amplitude $\lambda_{+}(k_{x_c})$, and by the Uhlmann geometric phase at critical momentum $k_{x_c}$\,\cite{Huang2014}:
\begin{equation}
    \Phi_{\text{U}}(k_{x_c})=\arg[\text{Tr}M(k_{x_c})]=\mu_{+}(k_{x_c}).
\end{equation}
Crucially, in the small-ramp-velocity limit $v\to 0$, we find that the Uhlmann geometric phase of the dephased state at $k_{x_c}$ reduces to the Berry geometric phase of the lower band of the final Hamiltonian (see Appendix\,\ref{appena4} for details):
\begin{equation}\label{eq.UC}
\Phi_{\text{U}}^{\rm de}(k_{x_c})=\arg\bigl[\cos(\Phi_\text{B}(k_{x_c})+\mathcal{O}(v_{\text{LZ}}^{1/2}))\bigr],
\end{equation}
where the Landau-Zener velocity $v_{\text{LZ}}=|m_f-m_c|v$. 
Accordingly, the dephased state $\rho^{\text{de}}_{\mathbf{k}}$ is topologically nontrivial if the $\Phi_{\text{U}}^{\rm de}(k_{x_c})=\mu_{+}(k_{x_c})=\pi$; otherwise it is topologically trivial.
Thus, for a small ramp velocity, the topological number of the dephased state is identical to that of the final Hamiltonian $H(m_f)$.

\begin{figure*}
  \centering
  \includegraphics[width=0.98\textwidth]{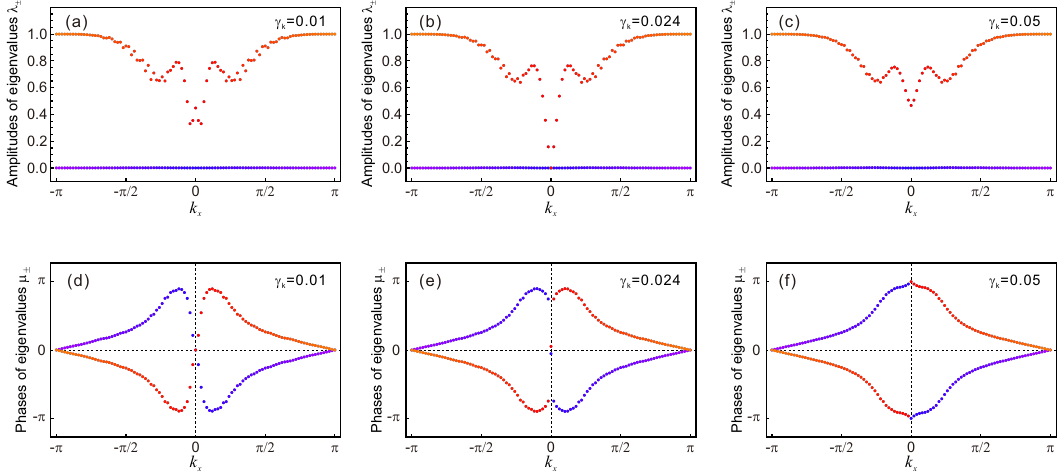}
  \caption{\textbf{Noise-induced topological phase transition}.
  Top panels (a-c) show the amplitude spectra of the holonomy eigenvalues $\lambda_{\pm}(k_x)$ of the dephased final state at $t=25$ for three dephasing rates $\gamma_{\mathbf{k}}=0.01$, $0.024$, and $0.05$, respectively.
  Bottom panels (d-f) display the corresponding phases of the holonomy eigenvalues $\mu_{\pm}(k_x)$ as a function of $k_x$ for the same dephasing rates.
  Parameters are $m_i=-4$, $m_f=-1$, and $v=0.2$ for all panels.}\label{fig.results3}
\end{figure*}

\section{\label{sec3}Results}
In this section, we numerically investigate the nonunitary preparation of the topologically nontrivial states and analyze the dependence on the dephasing rate and the parameter ramp velocity.

\subsection{Dynamical preparation of nontrivial states}
Here, we consider a protocol in which the Hamiltonian is ramped from a topologically trivial regime to a nontrivial one by varying the parameter $m(t)=m_i+(m_f-m_i)(1-e^{-vt})$ with ramp velocity $v=0.2$.
As illustrated in Fig.\,\ref{fig.results1}(a), the initial state $\rho(t=0)$ is prepared as the ground state of a trivial Hamiltonian (star symbol with $m_i=-4$), and the Hamiltonian parameter is subsequently ramped to a target nontrivial point (diamond symbol with $m_f=-1$), crossing a topological phase boundary at $m_c=-2$ that is accompanied by an energy-gap closing at the critical momentum $\mathbf{k}_c=0$.

As $m(t)$ is varied, the lower-band Chern number of the instantaneous Hamiltonian $H(m(t))$ changes from $0$ to $+1$ after the critical time, as indicated by the purple dashed line in Fig.\,\ref{fig.results1}(b).
However, for coherent evolution governed by
$\dot{\rho}_{\mathbf{k}}=-\mathrm{i}[H_{\mathbf{k}},\rho_{\mathbf{k}}]$, the Chern number of the evolved state (bar chart) remains invariant in time, since the coherent dynamics consists solely of local unitary transformations.
In Fig.\,\ref{fig.results1}(d), we numerically evaluate the Berry geometric phase—or equivalently, the Uhlmann geometric phase at zero temperature—of the coherently evolved final state at $t=25$ for different values of $k_x\in[-\pi,\pi]$ with equal spacing.
The geometric phase varies continuously, and the trivial state exhibits no winding as $k_x$ traverses the first Brillouin zone.
Thus, topologically nontrivial states cannot be reached from a trivial initial state through unitary dynamics alone.

To overcome this unitary limitation, we introduce dephasing noise, such that the evolution of the state is governed by the master equation $\overset{.}{\rho}_{\mathbf{k}} = -\text{i}[H_{\mathbf{k}},\rho_\mathbf{k}] + \gamma_{\mathbf{k}}[\tilde{\sigma}^z_\mathbf{k} \rho_\mathbf{k} \tilde{\sigma}^z_\mathbf{k} - \rho_\mathbf{k}]$.
The dephasing rate is chosen to be sufficiently large, $\gamma_{\mathbf{k}}=2$, ensuring that the coherence between the ground and excited states is completely suppressed, and the instantaneous state is well described by a diagonal ensemble in the eigenbasis of $H(m(t))$.
As shown in Fig.\,\ref{fig.results1}(c), the Uhlmann number of the evolved dephased state closely follows that of the instantaneous Hamiltonian, thereby resolving the topological mismatch inherent in unitary dynamics (see Fig.\,\ref{fig.results1}(b)).
In Fig.\,\ref{fig.results1}(e), we plot the Uhlmann geometric phase $\Phi_{\mathrm{U}}(k_x)$ of the dephased final state at $t=25$.
A $2\pi$-discontinuous jump from $-\pi$ to $\pi$ is observed at the critical momentum $k_{x_c}=0$, signaling a nontrivial winding.
By comparing Figs.\,\ref{fig.results1}(e) and \ref{fig.results1}(f), we find that the topological number of the dephased final state, $n_{\mathrm{U}}=+1$, matches that of the ground state of the final Hamiltonian, $n_{\mathrm{C}}=+1$, with only a small deviation in $\Phi_{\mathrm{U}}(k_x)$ confined to the vicinity of the gap-closing point $k_{x_c}=0$ due to residual excitations.
These numerical results demonstrate that topologically nontrivial states can be reached from an initially trivial regime through dephasing-induced nonunitary dynamics.
In the slow-ramp limit $v\to0$, the resulting nontrivial state can be regarded as a pseudoground state of the target nontrivial Hamiltonian, containing only a small density of excitations.

\subsection{Topological phase transition in parameter ramp}

At zero temperature, a change in the Chern number of the ground state is necessarily accompanied by a closing of the energy gap.
Here, we numerically demonstrate that a change in the Uhlmann number of the dephased mixed state is likewise accompanied by a closing of the amplitude gap in the spectrum of the holonomy matrix $M$.
In Fig.\,\ref{fig.results2}, the upper and lower panels show, respectively, the amplitudes and phases of the complex holonomy eigenvalues for the evolving dephased state at three representative times.
For the two-band model considered here, the holonomy matrix possesses two complex eigenvalues $z_{\pm}=\lambda_{\pm}e^{\mathrm{i}\mu_{\pm}}$.
At an early time ($t=5$ for example), before the instantaneous Hamiltonian $H(m(t))$ is ramped into the nontrivial regime, the instantaneous state remains close to the ground state with negligible excitations.
In this case, the amplitudes $\lambda_{\pm}(k_x)$ are gapped, and the phases $\mu_{\pm}(k_x)$ do not sweep the full interval $[-\pi,\pi]$ as $k_x$ traverses the Brillouin zone, implying a trivial index $\tilde{n}_{\mathrm{U}}=0$.
At the critical time $t_c\approx5.68$, the amplitude gap closes at $k_x=0$, coinciding with the energy-gap closing point, and the phases do not exhibit a $2\pi$ discontinuity.
After crossing the critical time, the amplitude spectrum becomes gapped again.
In this regime, $\tilde{n}_{\mathrm{U}}=+1$: the phase $\mu_{+}(k_x)$ ($\mu_{-}(k_x)$), associated with the larger (smaller) amplitude, decreases (increases) monotonically as a function of $k_x$, interrupted by a $2\pi$-discontinuous jump at $k_{x_c}=0$.
We note that $\tilde{n}_{\mathrm{U}}=-1$ corresponds to the opposite winding, where $\mu_{+}(k_x)$ increases monotonically and is interrupted by a jump from $\pi$ to $-\pi$ at $k_{x_c}=0$.

\begin{figure*}
  \centering
  \includegraphics[width=0.98\textwidth]{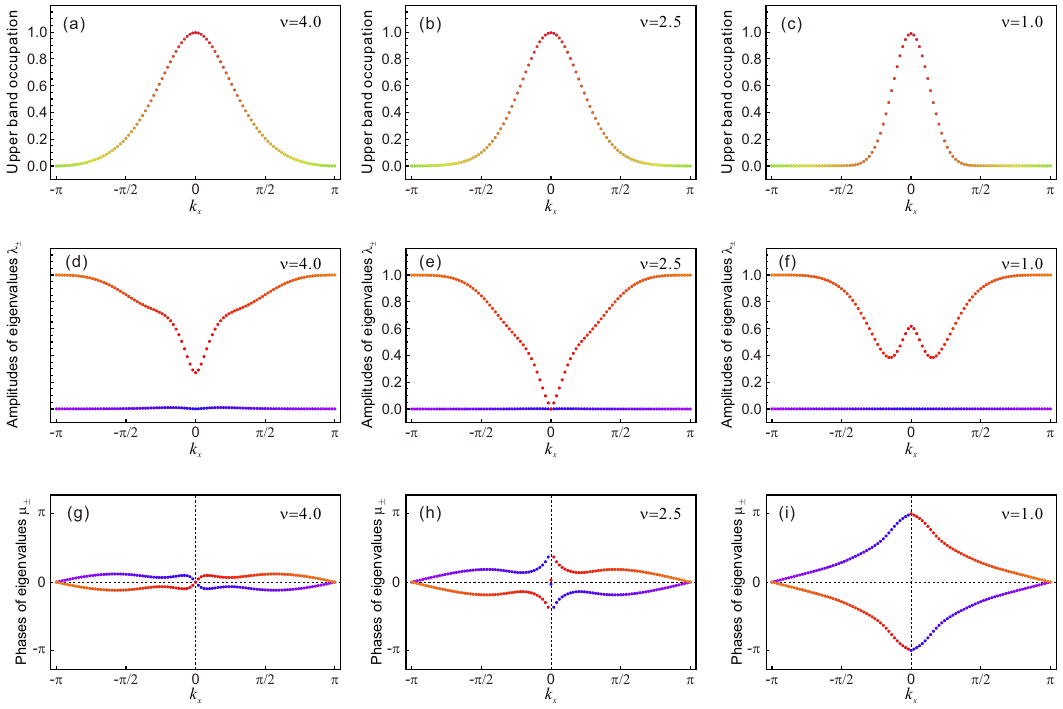}
  \caption{\textbf{Ramp-velocity-driven topological phase transition.}
  (a-c) Upper band occupation of the dephased final state at $t=25$, $[1+r_{k_{x},0}(t)]/2$, as a function of $k_x$ for three parameter ramp velocities $v=4.0$, $2.5$, and $1.0$, respectively.
  (d-f) Amplitude spectra of the holonomy eigenvalues $\lambda_{\pm}(k_x)$ of the dephased final state for the same parameter ramp velocities.
  (g-i) Phases of the holonomy eigenvalues $\mu_{\pm}(k_x)$ of the dephased final state for the same parameter ramp velocities.
  Parameters are $m_i=-4$, $m_f=-1$, and $\gamma_{\mathbf{k}}=2$ for all panels.}\label{fig.results4}
\end{figure*}

\subsection{Noise-induced topological phase transition}

In Fig.\,\ref{fig.results1}, we numerically showed that introducing finite dephasing noise can drive the topology of the final state from trivial to nontrivial.
This implies the existence of a critical noise strength that marks a topological phase transition in the dephased final state.
In Fig.\,\ref{fig.results3}, we plot the spectral flow of the complex holonomy eigenvalues $z_{\pm}(k_x)$ for the dephased final state $\rho^{\mathrm{de}}_{\mathbf{k}}(t=25)$ at three representative dephasing rates.
We identify three distinct noise regimes, characterized by qualitatively different spectral features:
(i) Weak-dephasing regime ($\gamma_{\mathbf{k}}=0.01$; panels (a) and (d)). 
The coherence of the final state is largely preserved, the amplitude spectra $\lambda_{\pm}(k_x)$ remain gapped, and the phase spectra $\mu_{\pm}(k_x)$ do not wind, corresponding to a trivial final state with topological index $\tilde{n}_{\mathrm{U}}=0$.
(ii) Intermediate-dephasing regime ($\gamma_{\mathbf{k}}=0.024$; panels (b) and (e)).
The amplitude gap closes at $k_{x_c}=0$, and the phase spectra also do not display a $2\pi$ discontinuity, indicating a topological transition point.
(iii) Strong-dephasing regime ($\gamma_{\mathbf{k}}=0.05$; panels (c) and (f)).
The coherence in the final state is largely suppressed, the amplitude gap reopens, and the phase spectra $\mu_{\pm}(k_x)$ exhibit a $2\pi$-discontinuous jump at the critical momentum $k_{x_c}=0$, signaling a nontrivial topology with index $\tilde{n}_{\mathrm{U}}=+1$.
Thus, increasing the dephasing rate modifies the spectrum structure of the holonomy eigenvalues and induces a topological phase transition, enabling the dephased final state to evolve from a topologically trivial state into a nontrivial one.

\subsection{Effects of parameter ramp velocity}
When the Hamiltonian parameter $m(t)$ is ramped from a trivial to a nontrivial regime, excitations are generated near the energy gap-closing point.
The resulting excitation density depends on the parameter ramp velocity $v$, as described by Landau–Zener physics (see Appendix\,\ref{appa1} for details).
In Fig.\,\ref{fig.results4}, we plot the excitation probability of the dephased final state at $k_y=0$, $P_{k_x,0}=[1+r_{k_{x},0}]/2$ [panels (a)-(c)], together with the corresponding spectral flow of the holonomy eigenvalues $z_{\pm}(k_x)$ [panels (d)-(i)] for three representative parameter ramp velocities.
We also identify three distinct ramp regimes.
(i) Large-ramp-velocity regime ($v=4.0$), excitations are generated throughout nearly the entire Brillouin zone, as shown in panel (a).
As a consequence, the phases $\mu_{\pm}(k_x)$ become continuous functions of $k_x$ while the amplitudes $\lambda_{\pm}(k_x)$ are gapped, indicating a topologically trivial final state with $\tilde{n}_{\mathrm{U}}=0$.
This behavior can be understood from Eq.\,(\ref{eq.UC}): the correction term $\mathcal{O}(v^{1/2}_{\text{LZ}})$ becomes sufficiently large to change the sign of $\text{Tr}M(k_{x_c})=\cos[\Phi_\text{B}(k_{x_c})+\mathcal{O}(v^{1/2}_{\text{LZ}})]$.
Consequently, the critical phase $\mu_{+}(k_{x_c})=\Phi_{\text{U}}(k_{x_c})$ jumps to zero and no longer coincides with the Berry geometric phase of the ground state of $H(m_f)$ at critical momentum.
(ii) In the intermediate-ramp-velocity regime ($v=2.5$), the amplitude spectra become gapless, and the phase spectra exhibit no winding, signaling the topological transition.
(iii) For the small-ramp-velocity regime ($v=1.0$), the excitation vicinity near the gap-closing point $k_{x_c}=0$ shrinks (panel (c)), and the correction term $\mathcal{O}(v^{1/2}_{\text{LZ}})$ is insufficient to change the sign of $\text{Tr}M(k_{x_c})$.
As a result, the phase spectra exhibit discontinuous $2\pi$ jumps, characteristic of a nontrivial topology.
Overall, increasing the parameter ramp velocity results in a higher excitation density and a pronounced deformation of the phase spectra relative to the ground-state result shown in Fig.\,\ref{fig.results1}(f), eventually driving the nontrivial dephased state to be trivial.

\section{\label{sec4}Discussion}
In this work, we have demonstrated a feasible protocol for preparing topologically nontrivial states from trivial regimes in 2D Qi-Wu-Zhang Chern insulators.
While a topological mismatch between the evolving state and the instantaneous Hamiltonian generally arises in coherent dynamics crossing nonadiabatic topological phase boundaries, we show that this mismatch can be reconciled by introducing dephasing noise, which fully suppresses the coherence between the ground and excited states.
As a result, in the limit of slow, nearly adiabatic nonunitary dynamics, one can reach a topologically nontrivial state with a nonzero topological number that coincides with the lower-band Chern number of the target nontrivial Hamiltonian.

Our results can be readily extended to other 2D topological insulators with higher Chern numbers\,\cite{Sticlet2012} and are of immediate relevance to current experiments.
Dephasing noise and the associated nonunitary dynamics can be experimentally realized via ensemble-averaged dynamics induced by disorder in the Hamiltonian spectrum\,\cite{Kropf2016,Gneiting2017,Chen2019}.
Furthermore, mixed-state geometric phases are experimentally accessible through quantum state tomography\,\cite{Wang2025,Wang2024} and interferometric schemes employing ancillary systems for purification\,\cite{Zhu2011,Kiselev2018,Viyuela2018,Mastandrea2026}.
These establish a realistic pathway for nonunitary preparation of nontrivial states and experimental measurement of topological Uhlmann invariants.

\section*{Acknowledgments}
We thank Bing Chen for the helpful discussions. This work was supported by Quantum Science and Technology-National Science and Technology Major Project (No.\,2021ZD0301200), National Natural Science Foundation of China (Nos.\,12474494, 12204468), Fundamental Research Funds for the Central Universities (No.\,WK2030000081), and China Postdoctoral Science Foundation (Nos.\,2024M753083, BX20240353). 

\appendix
\setcounter{equation}{0}
\renewcommand{\theequation}{A\arabic{equation}}

\section{\label{appendixa}Analytical derivation of quantized geometric phase at critical momenta for slow ramp}

\subsection{\label{appa1}Landau-Zener excitation probability}

We consider the two-band Hamiltonian of the Qi-Wu-Zhang Chern insulator:
\begin{equation}\nonumber
H_{\mathbf{k}}(m)=\sin k_x\sigma_x + \sin k_y\sigma_y + (m+\cos k_x+\cos k_y)\sigma_z,
\end{equation}
where $\sigma_i\,(i=x,y,z)$ are the Pauli matrices.
The energy band gaps close at the critical momenta $\mathbf{k_c}\in\{(0,0),\,(0,\pi),\,(\pi,0),\,(\pi,\pi)\}$, which occurs at the critical values of parameter $m_c\in\{0,\pm2\}$.
Here we focus on the gap-closing point $\mathbf{k}_c=(0,0)$ at the critical parameter $m_c=-2$ as a paradigmatic example.
The following analysis can be readily extended to the other critical momenta.

In the vicinity of $\mathbf{k}_c=(0,0)$, the Hamiltonian can be linearized and takes a Dirac form:
\begin{equation}
H_{\mathbf{k}}(m) = k_x\sigma_x+k_y\sigma_y+(m+2)\,\sigma_z.
\end{equation}
We ramp the Hamiltonian parameter via an exponential protocol:
\begin{equation}
m(t)=m_f-(m_f-m_i)e^{-vt},
\end{equation}
where $v>0$ controls the parameter ramp velocity.
When the ramp crosses the phase boundary with the critical value $m_c=-2$, the critical time $t_c$ is determined by:
\begin{equation}
t_c=\frac{1}{v}\ln\frac{m_f-m_i}{m_f+2}.
\end{equation}
The instantaneous slope at the critical time is:
\begin{equation}
\dot m(t)=v(m_f-m_i)e^{-vt}
\,\Rightarrow\,
\dot m(t_c)=v(m_f+2),
\end{equation}
so that, in the vicinity of $t_c$, the ramp is effectively linear.
Accordingly, the Landau-Zener velocity reads:
\begin{equation}
v_{\text{LZ}}=|\dot m(t_c)|=v\,|m_f+2|.
\end{equation}
For a slow parameter ramp across the gap-closing boundary, the probability $P_{\text{LZ}}^{\textbf{k}}$ of exciting from the lower to the upper energy band can be calculated as (setting $\hbar=1$):
\begin{equation}
P_{\rm LZ}^{\mathbf k} \approx e^{-\frac{\pi\Delta^2}{|\dot{m}(t_c)|}} = e^{-\frac{\pi\textbf{k}^2}{v_{\text{LZ}}}}.
\end{equation}

\subsection{Dephased state}

The system is assumed to evolve according to a master equation that incorporates pure dephasing noise in the eigenbasis of the instantaneous Hamiltonian:
\begin{equation}
\dot{\rho}_{\mathbf{k}} = -\mathrm{i}[H_{\mathbf{k}},\rho_{\mathbf{k}}] + \gamma_{\mathbf{k}} [\tilde{\sigma}^{z}_{\mathbf{k}}\rho_{\mathbf{k}}\tilde{\sigma}^{z}_{\mathbf{k}} - \rho_{\mathbf{k}}],
\end{equation}
where the operator $\tilde{\sigma}^z_\mathbf{k} (t) = \mathbf{n}_{\mathbf{k}}(t)\cdot\boldsymbol{\sigma}$
is the Pauli operator defined in the instantaneous eigenbasis of $H_{\mathbf{k}}(m(t))$.
This pure dephasing term suppresses coherence between the excited and ground states while leaving the band populations unchanged.
In the strong-dephasing regime for all times during the parameter ramp, the off-diagonal density-matrix elements decay rapidly in the instantaneous eigenbasis.
As a result, the dephased state at each momentum ${\mathbf{k}}$ takes the diagonal form:
\begin{equation}\label{eq.de}
\rho^{\text{de}}_{\mathbf{k}} = \frac{1}{2}[\mathbb{1}+r_\mathbf k\tilde{\sigma}^z_\mathbf{k}],
\end{equation}
where $r_\mathbf k=2P_{\rm LZ}^{\mathbf k}-1\in[-1,1]$.

\subsection{\texorpdfstring{Uhlmann geometric phase of dephased state at critical momentum $k_{x_c}$}{Uhlmann geometric phase of dephased state at critical momentum kxc}}

The Uhlmann connection associated with the dephased final state is defined as:
\begin{equation}
A_{\mathrm{U}} = \sum_{u,i,j} \frac{ \bra{n_{i}} [\partial_{k_{u}}\sqrt{\rho_{\mathbf{k}}^{\text{de}}},\sqrt{\rho_{\mathbf{k}}^{\text{de}}}] \ket{n_{j}} }{ p_{i}+p_{j} } \ket{n_{i}}\bra{n_{j}}\,dk_{u},
\end{equation}
where $\{|n_{\pm}(\mathbf k)\rangle\}$ are the eigenstates of the final Hamiltonian $H(m_f)$, and
$\rho_{\mathbf{k}}^{\mathrm{de}}=\sum_{i=\pm}p_i(\mathbf k)\,|n_i(\mathbf k)\rangle\langle n_i(\mathbf k)|$.
Substituting Eq.\,\eqref{eq.de} into it, and evaluating the commutator, one obtains the compact $2\times2$ form:
\begin{equation}
A_{\mathrm{U}}(k_y) = i\,\frac{N^2_\mathbf k}{2}\,(\partial_{k_y} \mathbf{n}_{\mathbf{k}}\times\mathbf{n}_{\mathbf{k}})\cdot\boldsymbol\sigma dk_y,
\end{equation}
where $\partial_{k_y}$ denotes the derivative along the $k_y$ loop, $N_\mathbf k=(\sqrt{1+r_\mathbf k}-\sqrt{1-r_\mathbf k})/\sqrt{2}$ is a real function, and $\mathbf n_{\mathbf k}$ is the unit Bloch vector of the final Hamiltonian.

At the gap-closing point $k_{x_c}=0$, the final Hamiltonian can be written as:
\begin{equation}
H_f(0,k_y)=\sin k_y\,\sigma_y+\bigl(m_f+1+\cos k_y\bigr)\sigma_z,
\end{equation}
so that the unit vector $\mathbf{n}(0,k_y)$ lies in the $y\sminus z$ plane.
Consequently, the vector $(\partial_{k_y}\mathbf n_{\mathbf{k}}\times \mathbf n_{\mathbf{k}})$ always points along $\hat{\mathbf x}$, i.e.
\begin{equation}
(\partial_{k_y}\mathbf{n}_{\mathbf{k}}\times \mathbf{n}_{\mathbf{k}})\cdot\boldsymbol\sigma = \bigl(\partial_{k_y}\mathbf n_{\mathbf{k}}\times \mathbf n_{\mathbf{k}}\bigr)_x\,\sigma_x.
\end{equation}
Introduce the planar angle:
\begin{equation}
\alpha(k_y)= \operatorname{atan2}\!\bigl(\sin k_y,\;m_f+1+\cos k_y\bigr),
\end{equation}
where the function $\operatorname{atan2}$ is the 2-argument arctangent.
Because $\mathbf n_{\mathbf{k}}$ lies in the $y\sminus z$ plane, one can then parameterize it as:
\begin{equation}
\mathbf n(0,k_y)=(0,\ \sin\alpha(k_y),\ \cos\alpha(k_y)),
\end{equation}
which implies the identity:
\begin{equation}
\bigl(\partial_{k_y}\mathbf{n}_{\mathbf{k}}\times \mathbf{n}_{\mathbf{k}}\bigr)_x=\partial_{k_y}\alpha(k_y).
\end{equation}
Moreover, using $\frac{d}{dk}\operatorname{atan2}(y,x)=\frac{x y'-y x'}{x^{2}+y^{2}}$ with
$y=\sin k_y$ and $x=m_f+1+\cos k_y$, one finds the explicit real expression:
\begin{equation}
\partial_{k_y}\alpha(k_y) = \frac{1+(m_f+1)\cos k_y}{1+(m_f+1)^2+2(m_f+1)\cos k_y}.
\label{eq.nkx}
\end{equation}

The Uhlmann geometric phase for a closed $k_y$ loop at the point $k_x$ is defined by:
\begin{equation}
\Phi_{\mathrm U}(k_{x}) = \arg\!\left(
\mathrm{Tr}\Bigl[\rho_{\mathbf k(0)}^{\mathrm{de}}\,
\mathcal P e^{\oint_{k_y} A_{\mathrm U}}\Bigr]
\right),
\end{equation}
where $\mathbf k(0)=(k_x,k_{y_0})$ specifies the initial point of the $k_y$ loop at $k_x$ and $\mathcal P$ denotes path ordering.
The holonomy matrix takes the form:
\begin{equation}\label{eq.holo}
M(k_x) = \rho_{\mathbf k(0)}^{\mathrm{de}}\,\mathcal P e^{\oint_{k_y} A_{\mathrm U}}.
\end{equation}
At the gap-closing point $k_{x_c}$, the Uhlmann connection reads:
\begin{equation}
A_{\mathrm U}(0,k_y)=\mathrm{i}\,\frac{N^2(0,k_y)}{2}\,\partial_{k_y}\alpha(k_y)\,\sigma_x\,dk_y.
\end{equation}
Since $A_{\mathrm U}(0,k_y)$ is proportional to $\sigma_x$ for all $k_y$, the connection matrices commute $[A_{\mathrm U}(0,k_y),A_{\mathrm U}(0,k_y')]=0$ at different $k_y$, and the path ordering becomes:
\begin{equation}
    \mathcal P e^{\oint_{k_y} A_{\mathrm U}(0,k_y)}=e^{i\Theta_0\sigma_x} = \cos \Theta_0 \mathbb{1}+i\sin \Theta_0\sigma_x,
\end{equation}
with 
\begin{equation}\label{eq.TrM}
\Theta_0 = \int_{-\pi}^{\pi} dk_y\;\frac{N^2(0,k_y)}{2}\,\partial_{k_y}\alpha(k_y).
\end{equation}
Substituting these into Eq.\,\eqref{eq.holo} gives the holonomy matrix at $k_{x_c}=0$:
\begin{equation}
M(k_{x_c})
=
\begin{pmatrix}
p_+^{(0)}\cos \Theta_0 & \mathrm{i}\,p_+^{(0)}\sin \Theta_0\\[3pt]
\mathrm{i}\,p_-^{(0)}\sin \Theta_0 & p_-^{(0)}\cos \Theta_0
\end{pmatrix},
\end{equation}
with the two eigenvalues of the holonomy matrix:
\begin{equation}
z_\pm(k_{x_c})=\frac12\Bigl(\cos \Theta_0 \pm \sqrt{r_0^2-\sin^2\Theta_0}\Bigr).
\end{equation}
Here $\sum_{i=\pm} p_i^{(0)} = \sum_{i=\pm}p_i(k_{x_c},k_{y_0})=1$ and $r_0 = p_+^{(0)}-p_-^{(0)}$.
Thus, the Uhlmann geometric phase at the gap-closing point $k_{x_c}=0$ can be calculated and written as:
\begin{equation}
\Phi_{\text{U}}^{\rm de}(k_{x_c})=\arg[\text{Tr}M(k_{x_c})]=\arg\bigl[\cos\Theta_0\bigr].
\end{equation}

\subsection{\texorpdfstring
  {Relation between Uhlmann and Berry phases at critical momentum $k_{x_c}$}
  {Relation between Uhlmann and Berry phases at critical momentum kxc}
\label{appena4}}

Along the $k_y$ loop at $k_{x_c}=0$, we consider a single critical point at $k_{y_c}=0$ for $m_c=-2$.
For a sufficiently slow ramp $v\to0$, the Landau-Zener excitations are appreciable only in a narrow neighborhood of the gap-closing point $(k_{x_c},k_{y_c})=(0,0)$.
We therefore introduce a small cutoff $k_a$ and decompose the Brillouin-zone integral into the narrow region $R_2$ and the other regular region $R_1$:
\begin{equation}
R_2= \{k_y:\ |k_y|\le k_a\},\quad R_1= [-\pi,\pi]\setminus R_2,
\end{equation}
where we choose $k_a=\Lambda\sqrt{v_{\mathrm{LZ}}}$ with $\Lambda\ge 1$.
We split Eq.\,\eqref{eq.TrM} as:
\begin{equation}\nonumber
\Theta_0= \int_{R_1}\! dk_y\;\frac{N^2(0,k_y)}{2}\,\partial_{k_y}\alpha + \int_{R_2}\! dk_y\;\frac{N^2(0,k_y)}{2}\,\partial_{k_y}\alpha.
\end{equation}
In the regular region $R_1$, the excitation probability $P_{\mathrm{LZ}}(0,k_y)$ is exponentially small, so that $|r(0,k_y)|\simeq 1$ and hence $N^2(0,k_y)/2\simeq 1/2$. This yields:
\begin{equation}\nonumber
\Theta_0\simeq \int_{R_1}\! dk_y\;\frac12\,\partial_{k_y}\alpha(k_y) + \int_{R_2}\! dk_y\;\frac{N^2(0,k_y)}{2}\,\partial_{k_y}\alpha(k_y).
\end{equation}
We add and subtract the $\frac12$-weight contribution over the narrow region $R_2$, and then:
\begin{align}
\Theta_0 &\simeq
\frac12\oint dk_y\;\partial_{k_y}\alpha(k_y) + \int_{R_2}\! dk_y\frac{N^2(0,k_y)-1}{2}\partial_{k_y}\alpha(k_y)
\nonumber\\ &= \Phi_{\text{B}}(k_{x_c}) + \delta\Phi,
\end{align}
where $\Phi_{\text{B}}(k_{x_c})$ is the Berry geometric phase of the lower band of the final Hamiltonian $H(m_f)$ at the critical momentum $k_{x_c}=0$ and is quantized to $\pi$ for $0<|m_f|<2$ and $0$ for otherwise.

The correction term $\delta\Phi$ is controlled by the vicinity of $k_{y_c}=0$:
\begin{equation}
\delta\Phi = \int_{R_2}\! dk_y\frac{N^2(0,k_y)-1}{2}\partial_{k_y}\alpha(k_y),
\end{equation}
Using $N=(\sqrt{1+r}-\sqrt{1-r})/\sqrt{2}$ and $r=2P_{\mathrm{LZ}}-1$, one has $(N^2-1)/2=-\sqrt{P_{\mathrm{LZ}}(1-P_{\mathrm{LZ}})}$.
Therefore, the correction term can be written as:
\begin{equation}\nonumber
\delta\Phi = -\int_{-k_a}^{k_a}\! dk_y\; \sqrt{P_{\mathrm{LZ}}(0,k_y)\bigl[1-P_{\mathrm{LZ}}(0,k_y)\bigr]}\;\partial_{k_y}\alpha(k_y).
\end{equation}
Expanding the factor $\sqrt{P_{\mathrm{LZ}}(1-P_{\mathrm{LZ}})}$ near $k_y=0$ yields:
\begin{equation}\nonumber
\sqrt{P_{\mathrm{LZ}}(0,k_y)\bigl[1-P_{\mathrm{LZ}}(0,k_y)\bigr]} = \sqrt{\frac{\pi}{v_{\mathrm{LZ}}}}\,|k_y| +\mathcal O\!\left(|k_y|^3/{v_{\mathrm{LZ}}^{3/2}}\right).
\end{equation}
On the other hand, for $m_f\neq -2$, $\partial_{k_y}\alpha(k_y)$ is analytic at $k_y=0$ (see Eq.\,\eqref{eq.nkx}):
\begin{equation}
\partial_{k_y}\alpha(k_y) = \partial_{k_y}\alpha(0)+\mathcal O(k_y^2) = \frac{1}{m_f+2}+\mathcal O(k_y^2).
\end{equation}
Substituting these into the correction term gives:
\begin{equation}
    \begin{split}
      \delta\Phi &\approx -\frac{1}{m_f+2}\sqrt{\frac{\pi}{v_{\mathrm{LZ}}}} \int_{-k_a}^{k_a}\! dk_y\,|k_y| \\ 
      &= -\frac{1}{m_f+2}\sqrt{\frac{\pi}{v_{\mathrm{LZ}}}}\;k_a^{\,2} = -\frac{\Lambda^2\sqrt{\pi}}{m_f+2}\,\sqrt{v_{\mathrm{LZ}}}\\ 
      & = \mathcal O\!\left(\sqrt{v_{\mathrm{LZ}}}\right).  
    \end{split}
\end{equation}
Thus,
\begin{equation}
\Phi_{\text{U}}^{\rm de}(k_{x_c})=\arg[\cos(\Phi_{\text{B}}(k_{x_c})+\mathcal{O}(v_{\text{LZ}}^{1/2}))].
\end{equation}
Overall, for sufficiently slow ramp $v\to0$, the Uhlmann geometric phase of the dephased final state at the critical momentum $k_{x_c}$ is quantized: $\Phi_{\mathrm U}^{\mathrm{de}}(k_{x_c})=\pi$ when the final Hamiltonian is in the topologically nontrivial regimes, and $\Phi_{\mathrm U}^{\mathrm{de}}(k_{x_c})=0$ for topologically trivial final Hamiltonian.

\bibliographystyle{apsrev4-1PRX.bst}
\bibliography{PRAreference}

\end{document}